\documentclass{elsart}

\journal{J. Comput. Phys.}

\usepackage{natbib}
\usepackage{definitions}
\usepackage{localdefs}
\usepackage{amsmath}
\usepackage{amssymb}
\usepackage{graphics}
\usepackage{amsbsy}
\usepackage{bm}

\begin{document}
\begin{frontmatter}

\title{Far-field approximation for hydrodynamic interactions in
parallel-wall geometry}

\author{S. Bhattacharya},
\author{J. Blawzdziewicz\thanksref{e-mail}},
\and
\author{E. Wajnryb\thanksref{IPPT}}
\address{Department of Mechanical Engineering, Yale University, New
Haven, CT 06520-8286, USA}
\thanks[e-mail]{E-mail: jerzy.blawzdziewicz@yale.edu}
\thanks[IPPT]{On leave from IPPT Warsaw, Poland}

\begin{abstract}

A complete analysis is presented for the far-field creeping flow
produced by a multipolar force distribution in a fluid confined
between two parallel planar walls.  We show that at distances larger
than several wall separations the flow field assumes the Hele-Shaw
form, i.e., it is parallel to the walls and varies quadratically in
the transverse direction.  The associated pressure field is a
two-dimensional harmonic function that is characterized by the same
multipolar number $m$ as the original force multipole.  Using these
results we derive asymptotic expressions for the Green's matrix that
represents Stokes flow in the wall-bounded fluid in terms of a
multipolar spherical basis.  This Green's matrix plays a central role
in our recently proposed algorithm [Physica A xx, {\bf xxx} (2005)] for
evaluating many-body hydrodynamic interactions in a suspension of
spherical particles in the parallel-wall geometry.  Implementation of
our asymptotic expressions in this algorithm increases its efficiency
substantially because the numerically expensive evaluation of the
exact matrix elements is needed only for the neighboring particles.
Our asymptotic analysis will also be useful in developing hydrodynamic
algorithms for wall-bounded periodic systems and implementing
acceleration methods by using corresponding results for the
two-dimensional scalar potential.

\end{abstract}


\end{frontmatter}

\section{Introduction}
\label{Introduction}

Numerical and theoretical investigations of particle motion in
suspensions bounded by planar walls require efficient methods for
evaluating hydrodynamic interactions in these systems.  Examples of
phenomena where the hydrodynamic wall effects are important include
collective particle motion in quasi-bidimensional colloidal
suspensions
\cite*[][]{%
Acuna_Campa-Carbajal_Tinoco-Arauz_Lara-Medina_Noyola:1998,%
Pesche-Kollmann-Nagele:2001,%
Santana_Solano-Arauz_Lara:2002,%
Stancik-Hawkinson:2004,%
Cohen-Mason-Weitz:2004%
}, 
and conformation dynamics of a DNA molecule in a parallel-plate
microchannel \cite{Chen-Graham-de_Pablo-Randall-Gupta-Doyle:2004}.

Several methods for evaluating hydrodynamic interactions in
wall-bounded systems have been proposed.  In some studies, the flow
reflected from the walls was calculated numerically using either
boundary-integral \cite{Durlofsky-Brady:1989} or finite-difference
\cite{Chen-Graham-de_Pablo-Randall-Gupta-Doyle:2004} techniques.  In a
different approach \cite{Staben-Zinchenko-Davis:2003}, the exact
point-force solution for the flow between the walls
\cite{Liron-Mochon:1976} was used.  Wall effects were also included
using a multiple-reflection technique
\cite{Bhattacharya-Blawzdziewicz:2002}, and several approximation
methods were proposed
\cite{Nott-Brady:1994,Lobry-Ostrowsky:1996,Pesche-Nagele:2000}.  While
all of these methods have their merits, they also have some essential
disadvantages, such as a high numerical cost or an insufficient (in
many cases unknown) accuracy.

Recently we have derived
\cite{Bhattacharya-Blawzdziewicz-Wajnryb:2005,%
Bhattacharya-Blawzdziewicz-Wajnryb:2005a},
a novel algorithm for evaluating hydrodynamic friction matrix in a
wall-bounded suspension of spheres under creeping-flow conditions.%
\footnote{An algorithm based on similar ideas was also developed by
Jones \protect\cite{Jones:2004}.}
Our Cartesian-representation method relies on transformations between
a spherical basis set of solutions of Stokes equations (this set is
consistent with the particle geometry) and a Cartesian basis set
(which is consistent with the wall geometry).  The algorithm provides
highly accurate results for multiparticle friction and mobility
matrices.

Using our approach, we have obtained several interesting numerical
results.  In particular, we have shown that the friction matrix
undergoes a crossover from the quasi-three-dimensional to
quasi-two-dimensional form when the interparticle distance becomes
larger than the wall separation $H$.  We have also observed an
unusually large resistance coefficient for a long rigid chain of
spheres in transverse motion (with respect to the orientation of the
chain) in a narrow, wall-bounded space.  Since both these effects
involve flow on the length scale $l\gg H$, they are not captured by
the usual single-wall superposition approximation which does not
properly take the far-field flow into account (as demonstrated in
\cite{Bhattacharya-Blawzdziewicz-Wajnryb:2005}).

Large-scale studies of particle dynamics in the two-wall geometry
require efficient simulation algorithms.  In our approach
\cite{Bhattacharya-Blawzdziewicz-Wajnryb:2005,%
Bhattacharya-Blawzdziewicz-Wajnryb:2005a}
the most expensive part is evaluation of the Green's matrix
$\GreenWallTotElement$ in the multipolar representation.  This matrix
is a key quantity in our algorithm---its elements correspond to the
coefficients in the expansion of the hydrodynamic Green's tensor for
the wall-bounded system into multipolar basis fields.  The inverse of
the Green's matrix combined with the one-particle reflection matrices
yields the multi-particle hydrodynamic friction matrix.

In our algorithm 
\cite{Bhattacharya-Blawzdziewicz-Wajnryb:2005,%
Bhattacharya-Blawzdziewicz-Wajnryb:2005a}
the matrix $\GreenWallTotElement$ is expressed in terms of lateral
Fourier integrals with respect to the two-dimensional wave vector in a
plane parallel to the walls.  Evaluation of these integrals is
especially difficult for widely separated particles due to the
oscillatory character of the integrands.  In the present paper we
derive much simpler asymptotic formulas for the matrix
$\GreenWallTotElement$ in the far-field regime.  When the particle
separation is sufficiently large, these formulas can be used instead
of the Fourier integrals, resulting in a significant reduction of
numerical cost (and in other important simplifications).

Our analysis of the asymptotic form of the matrix
$\GreenWallTotElement$ relies on the observation that in the far-field
regime the velocity field in the space between the walls assumes a
simple Hele-Shaw (i.e. the lubrication) form.  Accordingly, the flow
field has only the lateral components and it varies quadratically
across the space between the walls.  Such a flow field is entirely
determined by the corresponding pressure field, which is a
two-dimensional harmonic function that depends only on the lateral
coordinates.  It follows that at large distances $\totalDistance\gg
H$, the full three-dimensional hydrodynamic problem is reduced to a
much simpler two-dimensional scalar problem for the pressure.

This paper is organized as follows.  Our
method 
\cite{Bhattacharya-Blawzdziewicz-Wajnryb:2005,%
Bhattacharya-Blawzdziewicz-Wajnryb:2005a}
for evaluating many-particle hydrodynamic interactions in the
parallel-wall geometry is summarized in Secs.\ \ref{Multiparticle
hydrodynamic interactions} and \ref{Force-multipole expansion}.
Section \ref{Multiparticle hydrodynamic interactions} recalls the
induced-force formulation of the problem, and Sec.\
\ref{Force-multipole expansion} summarizes the force-multipole
expansion method.  The main theoretical results of the present
analysis are given in Sec.\ \ref{Far-field approximation}, where the
Hele-Shaw approximation for the far-field flow is discussed, and
explicit expressions for Green's matrix $\GreenWallTotElement$ are
derived.  In Sec.\ \ref{Section on Numerical implementation} we
present some results of numerical calculations.  We show the
dependence of Green's matrix elements on the interparticle distance,
and we illustrate the role of their far-field behavior in the description
of hydrodynamic interactions in rigid arrays of spheres.  Concluding
remarks are given in Sec.\ \ref{conclusions}, and some technical
details are presented in the appendices.

\section{Multiparticle hydrodynamic interactions}
\label{Multiparticle hydrodynamic interactions}

\subsection{Hydrodynamic resistance}
\label{section on hydrodynamic resistance}

We consider the motion of $N$ spherical particles of the radius $a$,
which are suspended in a fluid of viscosity $\eta$, under
creeping-flow conditions.  The system is bounded by two planar
parallel walls at the positions $z=0$ and $z=H$, where $\br=(x,y,z)$
are the Cartesian coordinates.  The centers of particles $i=1\ldots N$
are at positions $\bR_i=(X_i,Y_i,Z_i)$, and the translational and
rotational particle velocities are $\bU_i$ and $\bOmega_i$.  The
external forces and torques acting on the particles are denoted by
$\totForce_i$ and $\totTorque_i$. It is assumed that the flow field
satisfies the no-slip boundary conditions on the particle surfaces and
the walls.

For a system of spheres undergoing translational and rotational
rigid-body motion with no external flow, the particle dynamics is
characterized by the resistance matrix
\begin{equation}
\label{resistance matrix}
\resistanceMatrix_{ij}=
\left[
  \begin{array}{cc}
    \resistanceMatrixTT_{ij}&\resistanceMatrixTR_{ij}\\
    \resistanceMatrixRT_{ij}&\resistanceMatrixRR_{ij}
  \end{array}
\right],
\qquad i,j=1,\ldots,N,
\end{equation}
defined by the linear relation
\begin{equation}
\label{resistance relation}
\left[
  \begin{array}{c}
\totForce_i\\
\totTorque_i
  \end{array}
\right]
=
\sum_{j=1}^N
\left[
  \begin{array}{cc}
    \resistanceMatrixTT_{ij}&\resistanceMatrixTR_{ij}\\
    \resistanceMatrixRT_{ij}&\resistanceMatrixRR_{ij}
  \end{array}
\right]
\bcdot
\left[
  \begin{array}{c}
     \bU_j\\
     \bOmega_j
  \end{array}
\right].
\end{equation}
The dot in the above equation denotes the matrix multiplication and
contraction of the Cartesian tensorial components of the resistance
matrix.  Our goal is to calculate the resistance matrix
$\resistanceMatrix$, or its inverse, the mobility matrix
$\mobilityMatrix$.  Our method
\cite{Bhattacharya-Blawzdziewicz-Wajnryb:2005,%
Bhattacharya-Blawzdziewicz-Wajnryb:2005a} 
for evaluating these quantities is outlined below.

\subsection{Induced-force formulation}
\label{Induced-force formulation}

The effect of the suspended particles on the surrounding fluid can be
described in terms of the induced-force distributions on the particle
surfaces.  These distributions can be written in a form
\begin{equation}
\label{induced forces}
\bF_i(\br)=\deltab{a}(\br-\bR_i)\bff_i(\br),
\end{equation}
where 
\begin{equation}
\label{delta b}
\deltab{a}(\br)=a^{-2}\delta(r-a).
\end{equation}
By the definition of the induced force
\cite{Cox-Brenner:1967,Mazur-Bedeaux:1974,Felderhof:1976b}, the flow
field
\begin{equation}
\label{flow field produced by induced forces}
\bv(\br)=\sum_{i=1}^N
  \int\bT(\br,\br')\bcdot\bF_i(\br')\diff\br'
\end{equation}
is identical to the velocity field in the presence of the moving
particles.  Here
\begin{equation}
\label{Green's function}
\bT(\br,\br')=\bT_0(\br-\br')+\bT'(\br,\br')
\end{equation}
is the Green's function for the Stokes flow in the presence of the
boundaries; the Green's function $\bT(\br,\br')$ is decomposed into
the Oseen tensor 
\begin{equation}
\label{Oseen tensor}
\bT_0(\br)=\frac{1}{8\pi\eta r}(\identityTensor+\hat\br\hat\br),
\end{equation}
and the part $\bT'(\br,\br')$ that describes the flow reflected from
the walls.  In Eq.\ \refeq{flow field produced by induced forces}
it is assumed that the particles move with given velocities, but no
external flow is imposed.

The resistance relation \refeq{resistance relation} is linked to the
induced-force distributions \refeq{induced forces} through the
expressions
\begin{equation}
\label{force and torque}
   \totForce_i=\int\bF_i(\br)\diff\br,
\qquad
   \totTorque_i=\int\br_i\btimes\bF_i(\br)\diff\br
\end{equation}
for the total force and torque, respectively.  To determine the
resistance matrix \refeq{resistance matrix} we thus need to evaluate
the induced forces \refeq{induced forces} for given translational and
angular velocities of the particles.

\subsection{Boundary-integral equations for the induced forces}
\label{section on boundary-integral equations for the induced forces}

For a system of particles moving with the translational and angular
velocities $\bU_i$ and $\bOmega_i$, the induced-force distribution
\refeq{induced forces} can be obtained from the boundary-integral
equation 
\begin{eqnarray}
\label{boundary-integral equation for induced-force density}
   [\bZ_i^{-1}\bF_i](\br)
      +\sum_{j=1}^N\int
         [\delta_{ij}\bT'(\br-\br')
         +(1-\delta_{ij})
\bT(\br-\br')]
      \bcdot\bF_j(\br')\diff\br'
   =\bv_i^{\rm rb}(\br),\hspace{-4em}&&\nonumber\\
\br\in S_i,&&
\end{eqnarray}
where
\begin{equation}
\label{rigid-body velocity of drop i}
\bv_i^{\rm rb}(\br)=\bU_i+\bOmega_i\btimes\br_i
\end{equation}
is the rigid-body velocity field associated with the particle motion,
and $S_i$ is the surface of particle $i$.  In the boundary-integral
equation \refeq{boundary-integral equation for induced-force density},
$\bZ_i$ denotes the one-particle scattering operator that describes
the response of an individual particle to an external flow in an
unbounded space.  This operator is defined by the linear relation
\begin{equation}
\label{definition of operator Z}
\bF_i=-\bZ_i(\incidentVelocity{i}-\bv_i^{\rm rb}),
\end{equation}
where $\incidentVelocity{i}$ is the velocity incident to particle $i$.
For specific particle models (e.g., rigid particles or drops),
explicit expressions for the operator $\bZ_i$ are known
\cite[][]{%
Jones-Schmitz:1988,%
Cichocki-Felderhof-Schmitz:1988,%
Blawzdziewicz-Wajnryb-Loewenberg:1999%
}.  

\section{Force-multipole expansion}
\label{Force-multipole expansion}

\subsection{Spherical basis fields}
\label{section on spherical basis fields}

As in a standard force-multipole approach
\cite[][]{%
Cichocki-Felderhof-Hinsen-Wajnryb-Blawzdziewicz:1994,%
Cichocki-Jones-Kutteh-Wajnryb:2000%
}
the boundary-integral equation \refeq{boundary-integral equation for
induced-force density} is transformed into a linear matrix equation by
projecting it onto a spherical basis of Stokes flow.  To this end we
use the reciprocal basis sets defined by Cichocki et al.\
\cite{Cichocki-Felderhof-Schmitz:1988}.  We introduce, however, a
slightly different normalization to exploit the full symmetry of the
problem.

The singular and nonsingular spherical basis solutions of Stokes
equations $\sphericalBasisM{lm\sigma}(\br)$ and
$\sphericalBasisP{lm\sigma}(\br)$ (with $l=1,2,\ldots$;
$m=-l,\ldots,l$; and $\sigma=0,1,2$) have the following separable form
in the spherical coordinates $\br=(r,\theta,\phi)$:
\begin{subequations}
\label{spherical basis v +-}
\begin{equation}
\label{spherical basis v -}
\sphericalBasisM{lm\sigma}(\br)
       =\sphericalBasisCoefM{lm\sigma}(\theta,\phi)r^{-(l+\sigma)},
\end{equation}
\begin{equation}
\label{spherical basis v +}
\sphericalBasisP{lm\sigma}(\br)
       =\sphericalBasisCoefP{lm\sigma}(\theta,\phi)r^{l+\sigma-1}.
\end{equation}
\end{subequations}
The coefficients $\sphericalBasisCoefM{lm\sigma}(\theta,\phi)$ and
$\sphericalBasisCoefP{lm\sigma}(\theta,\phi)$ are combinations of
vector spherical harmonics with angular order $l$ and azimuthal order
$m$.  This property and the $r$-dependence in equations
\refeq{spherical basis v +-} define the Stokes-flow fields
$\sphericalBasisPM{lm\sigma}(\br)$ up to normalization.

We use here a convenient normalization introduced in
\cite{Bhattacharya-Blawzdziewicz-Wajnryb:2005a}, which emphasizes
various symmetries of the problem.  Explicit expressions for the
functions $\sphericalBasisCoefPM{lm\sigma}$ in this normalization are
given in Appendix \ref{Spherical basis}.  We note that both in our
present and in the original normalization
\cite{Cichocki-Felderhof-Schmitz:1988}, the basis fields
$\sphericalBasisM{lm\sigma}$ satisfy the identity
\cite{Perkins-Jones:1991} \refstepcounter{equation}
$$
\label{expansion of Oseen tensor in spherical basis}
\eta\bT_0(\br-\br')=
\left\{
   \begin{array}{ll}
     \displaystyle\sum_{lm\sigma}
        \sphericalBasisM{lm\sigma}(\br)\sphericalBasisPcon{lm\sigma}(\br'),
        \qquad &r>r',\\&\\
     \displaystyle\sum_{lm\sigma}
        \sphericalBasisP{lm\sigma}(\br)\sphericalBasisMcon{lm\sigma}(\br'),
        \qquad &r<r',
   \end{array}
\right.
\eqno{(\theequation{\mathit{a},\mathit{b}})}
$$ 
where $\bT_0(\br-\br')$ is the Oseen tensor \refeq{Oseen tensor}.
Relation \refeq{expansion of Oseen tensor in spherical basis} assures
that the Lorentz reciprocal symmetry of Stokes flow is reflected in
the symmetry of the resulting matrix representation of the problem
\cite[][]{Cichocki-Jones-Kutteh-Wajnryb:2000}.

Following \cite{Cichocki-Felderhof-Schmitz:1988} we also introduce the
reciprocal basis fields $\reciprocalSphericalBasisPM{lm\sigma}(\br)$,
defined by the orthogonality relations of the form
\begin{equation}
\label{orthogonality relations for spherical reciprocal basis}
\langle\deltab{a}\reciprocalSphericalBasisPM{lm\sigma}\mid
    \sphericalBasisPM{l'm'\sigma'}\rangle
    =\delta_{ll'}\delta_{mm'}\delta_{\sigma\sigma'}.
\end{equation}
Here
\begin{equation}
\label{scalar product}
\langle\bA\mid\bB\rangle=\int \bA^*(\br)\boldsymbol{\cdot}\bB(\br)\diff\br
\end{equation}
is the inner product, the asterisk denotes the complex conjugate, and
$\deltab{a}$ is defined in Eq.\ \refeq{delta b}.  The reciprocal basis
fields and the bra--ket notation \refeq{scalar product} allows us to
conveniently represent expansions of Stokes flow fields into the
complete sets of nonsingular and singular basis fields
\refeq{spherical basis v +-}.  In particular, any Stokes flow
$\bu(\br)$ that is nonsingular in the neighborhood of a point
$\br=\bR_i$ has an expansion
\begin{equation}
\label{expansion of Stokes flow in nonsingular basis}
\bu(\br)
  =\sum_{lm\sigma}\sphericalBasisP{lm\sigma}(\br-\bR_i)
\langle
   \deltab{a}(i)\reciprocalSphericalBasisP{lm\sigma}(i)
\mid
   \bu
\rangle,
\end{equation}
where 
\begin{subequations}
\label{fields centered at i}
\begin{equation}
\label{fields w centered at i}
\reciprocalSphericalBasisP{lm\sigma}(i)\equiv
   \reciprocalSphericalBasisP{lm\sigma}(\br-\bR_i),
\end{equation}
\begin{equation}
\label{delat centered at i}
\deltab{a}(i)\equiv\deltab{a}(\br-\bR_i).
\end{equation}
\end{subequations}

\subsection{Matrix representation}
\label{Section on matrix representation}

The matrix representation of the boundary-integral equation
\refeq{boundary-integral equation for induced-force density} is
obtained using the multipolar expansion 
\begin{equation}
\label{induced force in terms of multipoles}
\bF_i(\br)
   =\sum_{lm\sigma}
      f_i(lm\sigma)
            \deltab{a}(\br-\bR_i)
         \reciprocalSphericalBasisP{lm\sigma}(\br-\bR_i)
\end{equation}
of the induced-force distributions \refeq{induced forces}. The
multipolar moments in the above expression are given by the projection
\begin{equation}
\label{matrix element for multipolar moment}
f_i(lm\sigma)=\langle\sphericalBasisP{lm\sigma}(i)\mid\bF_i\rangle,
\end{equation} 
according to the orthogonality condition \refeq{orthogonality
relations for spherical reciprocal basis}.  The definition
\refeq{matrix element for multipolar moment} of the multipolar
expansion is justified by the identity
\begin{equation}
\label{flow produced by force multipole}
   \sphericalBasisM{lm\sigma}(\br)=
   \eta\int \bT_0(\br-\br')\deltab{a}(\br')
      \reciprocalSphericalBasisP{lm\sigma}(\br')\diff\br',
\end{equation}
which follows from the representation \refeq{expansion of Oseen tensor
in spherical basis} of the Oseen tensor.  Equations \refeq{induced
force in terms of multipoles} and \refeq{flow produced by force
multipole} indicate that the multipolar moments $f_i(lm\sigma)$ are
identical (apart from the trivial factor $\eta$) to the expansion
coefficients of the flow field scattered by an isolated particle in
unbounded space into the singular basis fields
$\sphericalBasisM{lm\sigma}$.

To obtain a linear matrix equation for the set of force multipolar
moments $f_i(lm\sigma)$, representation \refeq{induced
force in terms of multipoles} is inserted into the boundary-integral
equation \refeq{boundary-integral equation for induced-force density},
and the resulting expression is expanded into the nonsingular basis
fields \refeq{spherical basis v +}, which yields
\begin{equation}
\label{induced force equations in matrix notation}
   \sum_{j=1}^N\sum_{l'm'\sigma'}
      \GrandMobilityElement_{ij}(lm\sigma\mid l'm'\sigma')
      f_j(l'm'\sigma')
      =c_i(lm\sigma).
\end{equation}
For a particle moving in a quiescent fluid, the coefficients
\begin{equation}
\label{rigid-body coefficients}
c_i(lm\sigma)
  =\langle\deltab{a}(i)\reciprocalSphericalBasisP{lm\sigma}(i)
\mid
\bv_i^{\rm rb}
\rangle
\end{equation}
on the right side are nonzero only for $l=1$ and $\sigma=0,1$.  The
matrix $\GrandMobilityElement$ in Eq.\ \refeq{induced force equations
in matrix notation} consists of three contributions corresponding to
the three terms on the left side of Eq.\ \refeq{boundary-integral
equation for induced-force density},
\begin{eqnarray}
\label{Grand Mobility matrix}
      \GrandMobilityElement_{ij}(lm\sigma\mid l'm'\sigma')
   =
      \delta_{ij}\delta_{ll'}\delta_{mm'}Z_i^{-1}(l;\sigma\mid\sigma')
   +
      \delta_{ij}\GreenWallElement_{ij}(lm\sigma\mid l'm'\sigma')
&&\nonumber\\
   +
      (1-\delta_{ij})\GreenWallTotElement_{ij}(lm\sigma\mid l'm'\sigma').&&
\end{eqnarray}
Using the bra--ket notation these contributions can be expressed in
the form
\begin{equation}
\label{expression for single particle scattering matrix}
Z_i^{-1}(l;\sigma\mid\sigma')=
   \langle\deltab{a}(i)\reciprocalSphericalBasisP{lm\sigma}(i)
\mid
   \bZ_i^{-1}
\mid
    \deltab{a}(i)\reciprocalSphericalBasisP{lm\sigma'}(i)\rangle,
\end{equation}
\begin{equation}
\label{elements of wall-correction operator}
\GreenWallElement_{ii}(lm\sigma\mid l'm'\sigma')
   =\langle\deltab{a}(i)\reciprocalSphericalBasisP{lm\sigma}(i)
\mid
   \bT'
\mid
   \deltab{a}(i)\reciprocalSphericalBasisP{l'm'\sigma'}(i)\rangle,
\end{equation}
and
\begin{equation}
\label{elements of total wall operator}
\GreenWallTotElement_{ij}(lm\sigma\mid l'm'\sigma')
   =\langle\deltab{a}(i)\reciprocalSphericalBasisP{lm\sigma}(i)
\mid
   \bT
\mid
   \deltab{a}(j)\reciprocalSphericalBasisP{l'm'\sigma'}(j)\rangle.
\end{equation}
The first term $Z_i^{-1}(l;\sigma\mid\sigma')$ is associated with the
one particle operator $\bZ_i^{-1}$ in equation
\refeq{boundary-integral equation for induced-force density}, and it
relates the force multipoles $f_i(l'm'\sigma')$ induced on particle $i$
to the coefficients in the expansion of the flow field incoming to
this particle into the nonsingular spherical basis fields
\refeq{spherical basis v +}.  By the spherical symmetry, this term is
diagonal in the indices $l$ and $m$ and is independent of $m$.  The
Green's matrices $\GreenWallElement_{ij}(lm\sigma\mid l'm'\sigma')$
and \mbox{$\GreenWallTotElement_{ij}(lm\sigma\mid l'm'\sigma')$} are
associated with the integral operators that involve the kernels
$\bT'(\br,\br')$ and $\bT(\br,\br')$.  Using the orthogonality
relations \refeq{orthogonality relations for spherical reciprocal
basis} one can show that the elements of these matrices correspond to
the expansion of the flow produced by a force multipole centered at
$\bR_j$ into the nonsingular basis \refeq{spherical basis v +}
centered at $\bR_i$. 

Explicit expressions for the single-particle reflection matrix
$Z_i^{-1}$ are well known
\cite{Cichocki-Felderhof-Schmitz:1988,Felderhof-Jones:1989}.
Quadrature formulas for the Green's matrix
\mbox{$\GreenWallTotElement_{ij}$} have been derived in our recent
publication \cite{Bhattacharya-Blawzdziewicz-Wajnryb:2005a}, where the
matrix elements $\GreenWallTotElement_{ij}(lm\sigma\mid l'm'\sigma)$
are represented as a combination of the free-space Green's matrix
\cite{Felderhof-Jones:1989,Cichocki-Jones-Kutteh-Wajnryb:2000} and the
wall contribution $\GreenWallElement_{ij}(lm\sigma\mid l'm'\sigma)$
that is given in a form of a Hankel transform of a product of
several simple matrices.  The Hankel transform arises from angular
integration of lateral Fourier modes of Stokes flow.

The many-particle resistance matrix \refeq{resistance matrix} can be
obtained by solving Eq.\ \refeq{induced force equations in matrix
notation} and projecting the induced force multipoles onto the total
force and torque \refeq{force and torque}.  Explicit expressions for
the resistance matrix in terms of the generalized friction matrix
$\GrandMobilityElement^{-1}$ are given in
\cite{Bhattacharya-Blawzdziewicz-Wajnryb:2005a}.  In numerical
applications, the system of linear equations \refeq{induced force
equations in matrix notation} is truncated at a given multipolar order
$l$, and the resulting approximate friction matrix is supplemented
with a lubrication correction (as described in
\cite{Bhattacharya-Blawzdziewicz-Wajnryb:2005a}).

\section{Far-field approximation}
\label{Far-field approximation}

Calculation of the exact matrix elements
\mbox{$\GreenWallTotElement_{ij}(lm\sigma\mid l'm'\sigma')$} by our
Cartesian-representation method
\cite{Bhattacharya-Blawzdziewicz-Wajnryb:2005a} requires numerical
evaluation of Hankel transforms that involve the Bessel functions
$\BesselJ_{m-m'}(k\LateralDistance_{ij})$.  Here $k$ is the
magnitude of the lateral wave vector, and
$\LateralDistance_{ij}=|\LateralVector_i-\LateralVector_j|$, where
$\LateralVector_i=(X_i,Y_i)$ denotes the lateral position of particle
$i$.  For large interparticle distances $\LateralDistance_{ij}$, the
factor $\BesselJ_{m-m'}(k\LateralDistance_{ij})$ undergoes rapid
oscillations as a function of $k$.  Thus, evaluation of the Fourier
integrals in the Hankel transforms is numerically expensive for such
configurations.

In the following sections we derive explicit asymptotic expressions
for the matrix elements $\GreenWallTotElement_{ij}(lm\sigma\mid
l'm'\sigma')$ at large interparticle distances
$\LateralDistance_{ij}\gg H$.  As we will see, these expressions have
a very simple form, and do not require evaluation of the Fourier
integrals.

\subsection{Hele-Shaw form of the far-field flow}
\label{Section on Hele-Shaw approximation for the far-field flow}

Our asymptotic analysis relies on the observation that in the
far-field regime the flow between two parallel walls assumes the
Hele-Shaw form.  Accordingly, the asymptotic pressure field $p=p^\as$
varies only in the lateral direction, and the associated flow field
has the lubrication form
\begin{equation}
\label{Hele-Shaw flow field}
\bu^\as(\br)=-\half\eta^{-1}z(H-z)\bnablaLat p^\as(\lateralVector),
\end{equation}
where 
\begin{equation}
\label{definition of lateral vector}
\br=\lateralVector+z\ez
\end{equation}
and $\bnablaLat$ is the two-dimensional gradient operator with respect
to the lateral position $\lateralVector=(x,y)$.  By the
incompressibility of the flow field \refeq{Hele-Shaw flow field}, the
pressure field $p^\as$ satisfies the two-dimensional Laplace's
equation
\begin{equation}
\label{two dimensional Laplace's equation for  pressure}
\nablaLat^2p^\as(\lateralVector)=0.
\end{equation}
The asymptotic expressions \refeq{Hele-Shaw flow field} and \refeq{two
dimensional Laplace's equation for pressure} can be obtained
\cite{Bhattacharya:2005} by expanding the boundary-value problem for
Stokes flow in the space between the walls in the small parameter
$H/\lateralDistance\ll1$, where $\lateralDistance$ is the distance
from the force distribution that generates the fluid motion.  Since
the velocity field \refeq{Hele-Shaw flow field} itself satisfies the
Stokes equations and boundary conditions exactly, one can show that
the higher-order terms in the asymptotic expansion vanish.  This
property indicates that the correction terms are subdominant
\cite{Bender-Orszag:1999}, which in turn suggests that the asymptotic
behavior \refeq{Hele-Shaw flow field} and \refeq{two dimensional
Laplace's equation for pressure} is approached exponentially.  This
conclusion is consistent with the direct analysis of the asymptotic
form of the Green's function in the space between the walls by Liron
and Mochon \cite{Liron-Mochon:1976} (see the discussion in Sec.\
\ref{section on Asymptotic form of the two-wall Green's tensor}
below).

\subsection{Asymptotic basis sets}
\label{Basis sets of Hele-Shaw flows}

To find the far-field form of the velocity field produced by
induced-force multipoles and to obtain the corresponding asymptotic
expressions for the elements of the Green's matrix
$\GreenWallTotElement_{ij}(lm\sigma\mid l'm'\sigma')$, it is
convenient to define appropriate basis sets of Hele-Shaw flow and
pressure fields.  The sets of singular and nonsingular pressures are
defined by the relation
\begin{equation}
\label{pressure in terms of solid harmonics}
\PressureBasisPM{m}(\lateralVector)=\eta\ScalarBasisPM{m}(\lateralVector),
\end{equation}
where $m=0,\pm1,\pm2,\ldots$, and 
\begin{subequations}
\label{Scalar basis}
\begin{equation}
\label{Scalar basis minus}
\ScalarBasisM{0}(\lateralVector)=-\ln\lateralDistance,\qquad
\ScalarBasisM{m}(\lateralVector)=\frac{1}{2|m|}
   \lateralDistance^{-|m|}\e^{\im m\phi}, \quad m\not=0,
\end{equation}
\begin{equation}
\label{Scalar basis plus}
\ScalarBasisP{m}(\lateralVector)=
   \lateralDistance^{|m|}\e^{\im m\phi}
\end{equation}
\end{subequations}
are the two-dimensional harmonic basis functions.  The associated
Hele-Shaw basis velocity fields are 
\begin{equation}
\label{Hele-Shaw basis velocity fields}
\HeleShawBasisPM{m}(\br)
   =-\half z(H-z)\bnablaLat\ScalarBasisPM{m}(\lateralVector),
\end{equation}
according to Eq.\ \refeq{Hele-Shaw flow field}.  

Below we list several useful relations for the harmonic functions
\refeq{Scalar basis}.  First, we have the diagonal representation for
the Green's function
\refstepcounter{equation}
$$
\label{diagonal representation for Green's function}
\ScalarBasisM{0}(\lateralVector-\lateralVector')=
\left\{
\begin{array}{ll}
\displaystyle\sum_{m=-\infty}^{\infty}
\ScalarBasisM{m}(\lateralVector)\ScalarBasisPcon{m}(\lateralVector'),\qquad&\lateralDistance>\lateralDistance',\\
&\\
\displaystyle\sum_{m=-\infty}^{\infty}
\ScalarBasisP{m}(\lateralVector)\ScalarBasisMcon{m}(\lateralVector'),\qquad&\lateralDistance<\lateralDistance',
\end{array}
\right.
\eqno{(\theequation{\mathit{a},\mathit{b}})}
$$ 
which is analogous to the representation \refeq{expansion of Oseen
tensor in spherical basis} of the Oseen tensor.  Next, we also have
the displacement theorem
\begin{equation}
\label{displacement of scalar flow fields}
\ScalarBasisM{m'}(\lateralVector-\LateralVector_j)
   =\sum_{m=-\infty}^{\infty}
\ScalarBasisP{m}(\lateralVector-\LateralVector_i)
\ScalarDisplacementElements{+-}
(\LateralVector_{ij};m\mid m'),
\end{equation}
where $\LateralVector_{ij}=\LateralVector_i-\LateralVector_j$,
and the displacement matrix is given by
\begin{equation}
\label{expression for scalar displacement matrix}
\ScalarDisplacementElements{+-}(\LateralVector;m\mid m')
   =\theta(-mm')(-1)^{m'}
\frac{(|m|+|m'|)!}{|m|!|m'|!}
       \ScalarBasisM{m'-m}(\LateralVector).
\end{equation}
We note that due to the presence of the Heaviside step function
\begin{equation}
\label{Heaviside}
\theta(x)=\left\{
\begin{array}{ll}
0,\qquad&x<0,\\
1,\qquad&x\ge0,
\end{array}\right.
\end{equation}
in Eq.\ \refeq{expression for scalar displacement matrix}, the scalar
fields with the same sign of the indices $m,m'\not=0$ do not couple in
the displacement relation \refeq{displacement of scalar flow fields}.
We also note that the matrix \refeq{expression for scalar displacement
matrix} satisfies the symmetry relation
\begin{equation}
\label{symmetry of cylindrical displacement matrix}
\ScalarDisplacementElements{+-}(\LateralVector;m\mid m')
=\ScalarDisplacementElements{+-\,*}(-\LateralVector;m'\mid m).
\end{equation}

As a direct consequence of the displacement theorem
\refeq{displacement of scalar flow fields} for the scalar pressure
fields, we have the corresponding displacement relation for the
Hele-Shaw basis flows \refeq{Hele-Shaw basis velocity fields}
\begin{equation}
\label{displacement theorem for Hele-Shaw fields}
\HeleShawBasisM{m'}(\br-\LateralVector_j)=
   \sumPrim{m=-\infty}{\infty}
      \HeleShawBasisP{m}(\br-\LateralVector_i)
      \ScalarDisplacementElements{+-}
         (\LateralVector_{ij};m\mid m').
\end{equation}
The term with $m=0$ in the above relation vanishes because
$\HeleShawBasisP{0}\equiv0$ according to Eqs.\ \refeq{Scalar basis
plus} and \refeq{Hele-Shaw basis velocity fields}.  The prime at the
summation sign has been introduced to emphasize that this term is
omitted.

In the following section we will derive a diagonal representation
(analogous to \refeq{expansion of Oseen tensor in spherical basis} and
\refeq{diagonal representation for Green's function}) for the
hydrodynamic Green's tensor describing the asymptotic far-field
response of the fluid confined between walls to a point force.

\subsection{Asymptotic Green's tensor}
\label{Asymptotic Green's tensor}

An explicit expression for the far-field flow produced by a point
force in the space between the walls has been derived by Liron and
Mochon \cite{Liron-Mochon:1976} (see also \cite{Hackborn:1990}).
According to their results, the far-field flow produced by a force
$\bF$ applied at the position $(0,0,z')$ can be expressed in the form
\begin{equation}
\label{Liron-Mochon formula}
\bu(\br)={\textstyle\frac{3}{2}}\pi^{-1}\eta^{-1} H^{-3}z(H-z)z'(H-z')
\bnablaLat\bnablaLat
(-\ln\lateralDistance)\bcdot\bF
+o(\e^{-\lateralDistance/H}).
\end{equation}
The above relation can also be obtained by a direct expansion of the
boundary-value problem in the small parameter $H/\lateralDistance$
\cite{Bhattacharya:2005}.

Relation \refeq{Liron-Mochon formula} indicates that the correction to
the far-field $O(\lateralDistance^{-2})$ asymptotic behavior of the
fluid velocity $\bu$ decays exponentially with $\lateralDistance$.
Moreover, the vertical component of the force $\bF$ does not
contribute to the $O(\lateralDistance^{-2})$ behavior.

Equation \refeq{Liron-Mochon formula} can be rephrased as an
expression for the asymptotic form $\WallGreenFunctionAs$ of the full
Green's function \refeq{Green's function}
\begin{equation}
\label{asymptotic form of wall Green's function}
\WallGreenFunctionAs(\br,\br')
   =-{\textstyle\frac{3}{2}}\pi^{-1}\eta^{-1} H^{-3}z(H-z)z'(H-z')
\bnablaLat\bnablaLat'
(-\ln|\lateralVector-\lateralVector'|),
\end{equation}
where $\br'=\brho'+z'\ez'$.  One of the gradient operators in the
above formula has been applied to the primed coordinates to emphasize
the Lorentz symmetry of the Green's tensor
\begin{equation}
\label{Lorentz symmetry of asymptotic Green's function}
\WallGreenFunctionAs(\br,\br')={\WallGreenFunctionAs}^\dagger(\br',\br)
\end{equation}
(where the dagger denotes the transpose of the tensor).  Due to the
translational invariance of the system in the lateral directions, the
Green's function \refeq{asymptotic form of wall Green's function}
satisfies the identity
\begin{equation}
\label{lateral invariance of asymptotic Green's function}
\WallGreenFunctionAs(\br-\LateralVector,\br'-\LateralVector)
=\WallGreenFunctionAs(\br,\br'),
\end{equation}
where the vector $\LateralVector$ has only lateral components.

Using Eqs.\ \refeq{Hele-Shaw basis velocity fields} and \refeq{diagonal
representation for Green's function} and noting that the Green's
function \refeq{asymptotic form of wall Green's function} is quadratic
both in primed and unprimed transverse variables, we find the relation 
\refstepcounter{equation}
$$
\label{diagonal representation for asymptotic hydrodynamic Green's function}
\WallGreenFunctionAs(\br,\br')=-\frac{6}{\pi\eta H^{3}}
\left\{
\begin{array}{ll}
\displaystyle\sumPrim{m=-\infty}{\infty}
\HeleShawBasisM{m}(\br)\HeleShawBasisPcon{m}(\br'),
   \quad&\lateralDistance>\lateralDistance',\\
&\\
\displaystyle\sumPrim{m=-\infty}{\infty}
\HeleShawBasisP{m}(\br)\HeleShawBasisMcon{m}(\br'),
   \quad&\lateralDistance<\lateralDistance',
\end{array}
\right.
\eqno{(\theequation{\mathit{a},\mathit{b}})}
$$
which is analogous to the diagonal representation of the Oseen tensor
\refeq{expansion of Oseen tensor in spherical basis}.  Equation
\refsubeq{diagonal representation for asymptotic hydrodynamic Green's
function}{a} combined with the displacement theorem for the Hele-Shaw
basis fields \refeq{displacement theorem for Hele-Shaw fields} and
identity \refeq{lateral invariance of asymptotic Green's function}
yields the symmetric representation of the asymptotic Green's tensor
\refeq{asymptotic form of wall Green's function}
\begin{equation}
\label{asymptotic Green's function in symmetric representation}
\WallGreenFunctionAs(\br,\br')=
-\frac{6}{\pi\eta H^{3}}
   \sumPrim{m=-\infty}{\infty}\sumPrim{m'=-\infty}{\infty}
    \HeleShawBasisP{m}(\br-\LateralVector_i)
      \ScalarDisplacementElements{+-}
         (\LateralVector_{ij};m\mid m')
    \HeleShawBasisPcon{m'}(\br'-\LateralVector_j).
\end{equation}

\subsection{Asymptotic form of the two-wall Green's matrix}
\label{section on Asymptotic form of the two-wall Green's tensor}

The asymptotic form of the matrix elements \refeq{elements of total
wall operator} can be obtained by projecting relation
\refeq{asymptotic Green's function in symmetric representation} onto
the reciprocal basis fields $\reciprocalSphericalBasisP{lm\sigma}$
centered at the points $\bR_i$ and $\bR_j$.  The resulting expression
involves the matrix elements
\begin{equation}
\label{matrix elements between spherical and Hele Shaw bases}
\langle
\deltab{a}(i)
   \reciprocalSphericalBasisP{lm\sigma}(i)
\mid
   \HeleShawBasisP{m'}(i)
\rangle
   =\delta_{mm'}\TransformationElementSAs(Z_i;lm\sigma),
\end{equation}
where $\HeleShawBasisP{m'}(i)\equiv
\mbox{$\HeleShawBasisP{m'}(\br-\LateralVector_i)$}$.  The elements
\refeq{matrix elements between spherical and Hele Shaw bases} are
diagonal in the azimuthal number $m$ by cylindrical symmetry, they
depend only on the vertical coordinate $Z_i$ of the point
$\bR_i=\LateralVector_i+Z_i\ez$, and they are real.  Using these
properties, the following asymptotic form of the wall Green's matrix
\refeq{elements of total wall operator} is obtained
\begin{equation}
\label{asymptotic Green's matrix elements}
\GreenWallTotElementAs_{ij}(lm\sigma\mid l'm'\sigma')
=-\frac{6}{\pi\eta H^3}
\TransformationElementSAs(Z_i;lm\sigma)
   \ScalarDisplacementElements{+-}
      (\LateralVector_{ij};m\mid m')
         \TransformationElementSAs(Z_j;l'm'\sigma').\qquad
\end{equation}
Due to the symmetric structure of the expression \refeq{asymptotic
Green's matrix elements} and the symmetry property \refeq{symmetry of
cylindrical displacement matrix} of the scalar displacement matrix
$\ScalarDisplacementElements{+-}$, the Lorentz symmetry
\begin{equation}
\label{Lorentz symmetry of asymptotic Green's matrix}
\GreenWallTotElementAs_{ij}(lm\sigma\mid l'm'\sigma')
=\GreenWallTotElementAsCon_{ji}(l'm'\sigma'\mid lm\sigma)
\end{equation}
is manifest.  We note that the presence of the Heaviside step function
in relation \refeq{expression for scalar displacement matrix} implies
that
\begin{equation}
\label{terms with m of different sign vanish in G}
\GreenWallTotElementAs_{ij}(lm\sigma\mid l'm'\sigma')=0
\quad\mbox{for}\quad mm'\ge0.
\end{equation}

The physical interpretation of the matrix $\TransformationElementSAs$
follows from the expression
\begin{equation}
\label{expansion of Hele-Shaw basis field into spherical basis}
\HeleShawBasisP{m}(\br-\LateralVector_i)
   =\sum_{l\sigma}\sphericalBasisP{lm\sigma}(\br-\bR_i)
      \TransformationElementSAs(Z_i;lm\sigma),
\end{equation}
which results from Eqs.\ \refeq{expansion of Stokes flow in
nonsingular basis} and \refeq{matrix elements between spherical and
Hele Shaw bases}.  The matrix $\TransformationElementSAs(Z;lm\sigma)$
thus describes the transformation from the representation of the flow
in terms of nonsingular Hele-Shaw basis
$\HeleShawBasisP{m}(\br-\LateralVector_i)$ centered at the lateral
position $\LateralVector_i$ to the spherical representation
\refeq{spherical basis v +} centered at $\bR_i$.

\subsection{Multipolar flow fields}
\label{Multipolar flow fields}

An alternative interpretation of the matrix
$\TransformationElementSAs$ is obtained by considering the far-field
flow
\begin{equation}
\label{asymptotic wall flow produced by force multipole}
\wallSphericalBasisAsM{lm\sigma}(\br-\LateralDistance_2;Z_2)=
   \int \bT^\as(\br,\br')
\deltab{a}(\br'-\bR_2)
      \reciprocalSphericalBasisP{lm\sigma}(\br'-\bR_2)\diff\br'
\end{equation}
produced in the space between the walls by the multipolar force
density
\begin{equation}
\label{force multipole producing far field flow}
\bF(\br')=\deltab{a}(\br'-\bR_2)
         \reciprocalSphericalBasisP{lm\sigma}(\br'-\bR_2)
\end{equation}
centered at $\bR_2$.  By inserting representation \refsubeq{diagonal
representation for asymptotic hydrodynamic Green's function}{a}
specified for the shifted asymptotic Green's function \refeq{lateral
invariance of asymptotic Green's function} with
$\LateralVector=\LateralVector_2$ into \refeq{asymptotic wall flow
produced by force multipole} and using definition \refeq{matrix
elements between spherical and Hele Shaw bases} of the matrix
$\TransformationElementSAs$ we find that
\begin{equation}
\label{expression for multipolar asymptotic field in Hele-Shaw basis}
\wallSphericalBasisAsM{lm\sigma}(\br-\LateralVector_2;Z_2)
=-\frac{6}{\pi\eta H^{3}}
\HeleShawBasisM{m}(\br-\LateralVector_2)
\TransformationElementSAs(Z_2;lm\sigma).
\end{equation}
Thus, the matrix element $\TransformationElementSAs(Z_2;lm\sigma)$
represents the amplitude of the Hele-Shaw basis field
$\HeleShawBasisM{m}$ in the far-field multipolar velocity
\refeq{asymptotic wall flow produced by force multipole}.  Only one
term contributes to this flow according to Eq.\ \refeq{expression for
multipolar asymptotic field in Hele-Shaw basis} because of the
cylindrical symmetry of the problem.

The asymptotic multipolar flow fields \refeq{asymptotic wall flow
produced by force multipole} can also be expressed in terms of the
matrix elements \refeq{asymptotic Green's matrix elements}.  To this
end the right side of Eq.\ \refeq{asymptotic wall flow produced by
force multipole} is expanded in the spherical basis fields
\refeq{spherical basis v +} with the help of identity \refeq{expansion
of Stokes flow in nonsingular basis}. The expansion yields the
relation
\begin{equation}
\label{expansion of asymptotic multipolar fields into spherical + basis}
\wallSphericalBasisAsM{lm\sigma}(\br-\LateralVector_2;Z_2)
   =\sum_{l'm'\sigma'}\sphericalBasisP{l'm'\sigma'}(\br-\bR_1)
   \GreenWallTotElementAs_{12}(l'm'\sigma'\mid lm\sigma),
\end{equation}
where $\GreenWallTotElementAs_{12}$ is given by Eq.\ \refeq{elements
of total wall operator} with the Green's function $\bT$ replaced with
$ \bT^\as$.  The above expression relates the asymptotic flow
$\wallSphericalBasisAsM{lm\sigma}$ centered at the position $\bR_2$
and the spherical basis fields centered at a different position
$\bR_1$.

We note that for each $m$ only several force multipoles \refeq{force
multipole producing far field flow} produce a nonzero far-field
velocity \refeq{asymptotic wall flow produced by force
multipole}. This behavior results from the properties of the matrix
$\TransformationElementSAs(Z_2;lm\sigma)$ that appearers in relation
\refeq{expression for multipolar asymptotic field in Hele-Shaw basis};
the form of this matrix is analyzed in Sec.\ \ref{Explicit expression}
below.  A further discussion of the multipolar fields in the space
between the walls is given in Appendix \ref{Matrix elements and
asymptotic flow}.

\subsection{Explicit expressions for the transformation matrix 
$\TransformationElementSAs$}
\label{Explicit expression}

A general structure of the matrix $\TransformationElementSAs$ can be
inferred using scaling arguments.  According to Eq.\ \refeq{spherical
basis v +} spherical basis fields
$\sphericalBasisP{lm\sigma}(\br-\bR_i)$ are homogeneous functions of
the order $l+\sigma-1$ of the relative-position vector
$\br_i=\br-\bR_i$.  Similarly, Eqs.\ \refeq{Scalar basis plus} and
\refeq{Hele-Shaw basis velocity fields} imply that the Hele-Shaw basis
fields $\HeleShawBasisP{m}(\br-\LateralVector_i)$ are combinations of
homogeneous functions of the order $|m|-1$, $|m|$, and $|m|+1$ of
$\br_i$.  Since the coefficients
$\TransformationElementSAs(Z_i;lm\sigma)$ are independent of $\br_i$,
relation \refeq{expansion of Hele-Shaw basis field into spherical
basis} implies that the non-zero elements of
$\TransformationElementSAs(Z_i;lm\sigma)$ satisfy the condition
\begin{equation}
\label{condition for nonzero elements of C matrix}
l+\sigma-|m|\le2.
\end{equation}  
A detailed analysis of relation \refeq{expansion of Hele-Shaw basis
field into spherical basis} reveals that the nonzero elements of the
matrix $\TransformationElementSAs$ can be written in the form
\cite{Bhattacharya:2005}
\begin{equation}
\label{nonzero elements of C matrix}
\TransformationElementSAs(Z;l\,\,\mbox{$\pm\mu$}\,\,\sigma)=
   \matrixElementBPM{l-\mu\,\,\sigma}(\mu;Z),\qquad\mu=|m|\ge1,
\end{equation}
where $\matrixElementBPM{\lambda\,\sigma}(\mu;Z)$
are the elements of the $3\times3$ matrix
\begin{equation}
\label{expression for matrix B}
\left\{
   \matrixElementBPM{\lambda\,\sigma}(\mu;Z)
\right\}_{\lambda,\sigma=0,1,2}
=\half A^{\pm}(\mu)
   \left[
       \begin{array}{ccc}
          -Z(H-Z)&\mp(H-2Z)&2\\
&&\\
          \displaystyle\frac{-\mu(H-2Z)}{(\mu+1)(2\mu+3)^{1/2}}
                    &\displaystyle\frac{\pm2\mu}{(\mu+1)(2\mu+3)^{1/2}}
                   &0\\
&&\\
          \displaystyle\frac{2\mu(\mu+1)^{1/2}}{(\mu+2)(2\mu+3)(\mu+5)^{1/2}}
                   &0&0
       \end{array}
    \right],
\end{equation}
with
\begin{equation}
\label{coefficient A}
A^{\pm}(\mu)
  =(\mp 2)^{\mu}\,\mu!\,\left[\frac{4\pi}{(2\mu+1)(2\mu)!}\right]^{1/2}.
\end{equation}
The range $\lambda=0,1,2$ of the index $\lambda=l-|m|$ in equation
\refeq{expression for matrix B} result from the conditions $|m|\le l$
and \refeq{condition for nonzero elements of C matrix}.  All other
elements of the matrix $\TransformationElementSAs$ vanish.

We close our theoretical considerations with a remark that the
asymptotic form $\GreenWallTotElementAs_{ij}$ of the Green's matrix
$\GreenWallTotElement_{ij}$ is approached exponentially for
$R_{ij}\to\infty$ because Liron-Mochon formula \refeq{Liron-Mochon
formula} is exponentially accurate at large distances.  As in relation
\refeq{Liron-Mochon formula}, the lengthscale for this approach is set
by the wall separation $H$.  The asymptotic expression
\refeq{asymptotic Green's matrix elements} should thus be very
accurate when the interparticle distance $R_{ij}$ is larger than
several wall separations $H$.  This conclusion is supported by our
numerical results discussed in the following section.

\section{Numerical examples}
\label{Section on Numerical implementation}

\subsection{Matrix elements}
\label{Matrix elements}

A typical behavior of the Green's matrix $\GreenWallTotElement_{ij}$
is illustrated in Figs.\ \ref{plots of asymptotic behavior of matrix
elements 2} and \ref{plots of asymptotic behavior of matrix elements
m}.  The results are shown for the matrix elements
\begin{equation}
\label{shown matrix elements}
\GreenWallTotElement_{12}(1\,\,\mbox{$-1$}\,0\mid l\,\mu\,\sigma)
=(-1)^\sigma
   \GreenWallTotElementCon_{12}(1\,1\,0\mid l\,\,\mbox{$-\mu$}\,\sigma)
\end{equation}
versus the lateral distance $\LateralDistance_{12}$ scaled by the wall
separation $H$ for several values of the parameters $l,\sigma$, and
$\mu>0$.  The elements \refeq{shown matrix elements} play a special
role in our theory because in the asymptotic regime
$\LateralDistance_{12}\gg H$ they are directly related to the
far-field multipolar flow \refeq{asymptotic wall flow produced by
force multipole} according to Eq.\ \refeq{asymptotic multipolar flow
in terms of matrix elements}.  In all examples, the vertical
coordinates of the points $(1)$ and $(2)$ are $Z_1=Z_2=\quarter H$.
For other configurations the matrix elements have a similar behavior.

We present our results in the form of the rescaled elements defined by
the relation
\begin{equation}
\label{rescaled elements of Green matrix}
\GreenWallTotElement_{12}
   (1\,\,\mbox{$\mp1$}\,0\mid l\,\,\mbox{$\pm\mu$}\,\sigma)
=H^{\mu-l-\sigma+1}
   \tilde\GreenWallTotElement_{12}
   (1\,\,\mbox{$\mp1$}\,0\mid l\,\,\mbox{$\pm\mu$}\,\sigma)
      \ScalarBasisM{\pm(\mu+1)}(\LateralVector_{12}).
\end{equation}
For those values of parameters $l$ and $\mu$ for which the matrix
elements \refeq{rescaled elements of Green matrix} do not vanish, the
factor $\ScalarBasisM{\pm(\mu+1)}(\LateralVector_{12}) \sim
\LateralDistance_{12}^{-(\mu+1)}$ corresponds to the far-field
behavior of $\GreenWallTotElementAs_{12}(1\,\,\mbox{$\mp1$}\,0\mid
l\,\,\mbox{$\pm\mu$}\,\sigma)$, according to Eqs.\ \refeq{expression
for scalar displacement matrix} and \refeq{asymptotic Green's matrix
elements}.  In the asymptotic regime $\LateralDistance_{12}\gg H$ the
rescaled elements
\mbox{$\tilde\GreenWallTotElementAs_{12}(1\,\,\mbox{$\mp1$}\,0\mid
l\,\,\mbox{$\pm\mu$}\,\sigma)$} depend only on the vertical
coordinates $Z_1$ and $Z_2$.  The nonzero asymptotic elements are
quadratic functions of the vertical coordinate $Z_1$, and they are at
most quadratic in $Z_2$ but can also be linear or constant in this
variable, as indicated by Eqs.\ \refeq{asymptotic Green's matrix
elements} and \refeq{expression for matrix B}.  The far-field flow
\refeq{asymptotic wall flow produced by force multipole} is related to
these elements by Eq.\ \refeq{form of far field flow}.

\begin{figure}
\begin{center}
\begin{picture}(300,360)(0,0)
\put(0,0){\scalebox{1}
      {\includegraphics{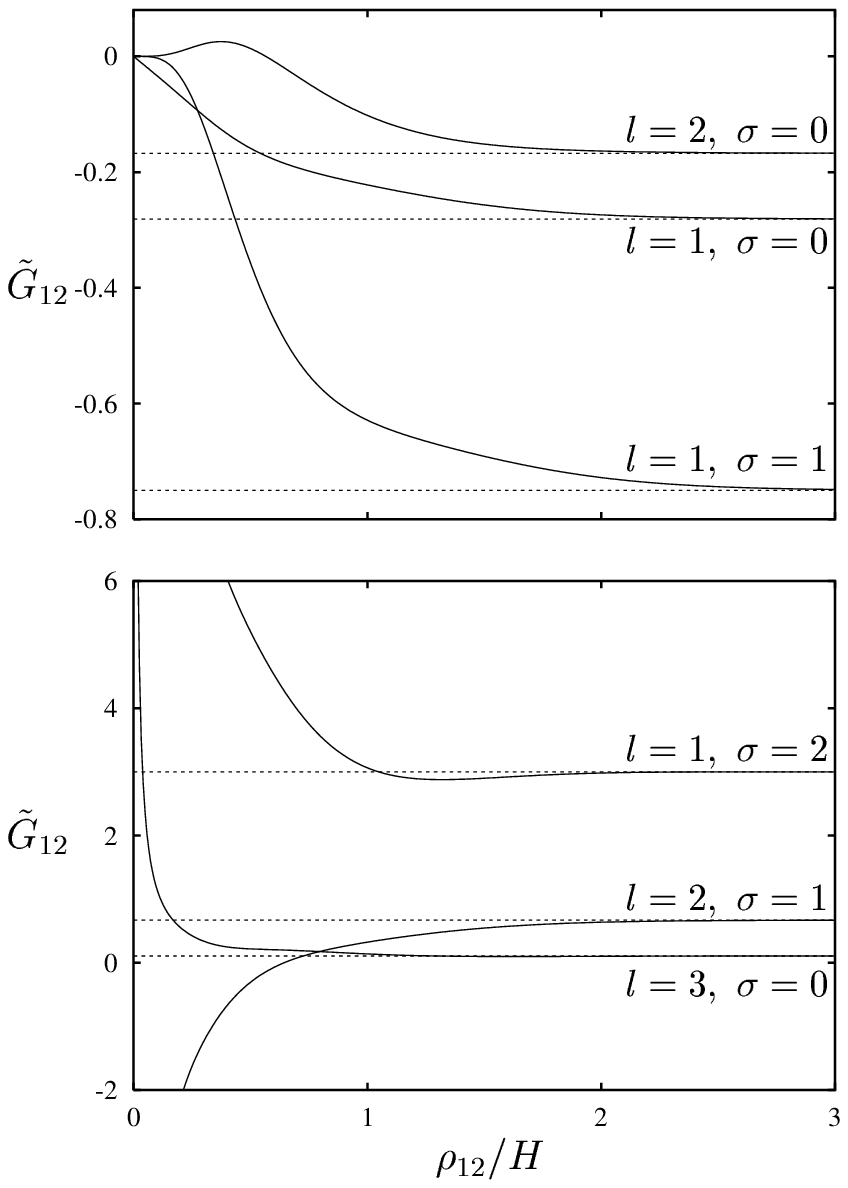}}}
\end{picture}
\end{center}
\caption{
Rescaled matrix elements
$\tilde\GreenWallTotElement_{12}(1\,\,\mbox{$-1$}\,0\mid
l\,1\,\sigma)$ versus lateral distance $\LateralDistance_{12}$ for
$Z_1=Z_2=\quarter H$; values of parameters $l$ and $\sigma$ as
labeled.  Exact solution (solid lines); asymptotic behavior (dotted
lines).
}
\label{plots of asymptotic behavior of matrix elements 2}
\end{figure}

\begin{figure}[t]
\begin{center}
\begin{picture}(300,180)(0,0)
\put(0,0){\scalebox{1}
      {\includegraphics{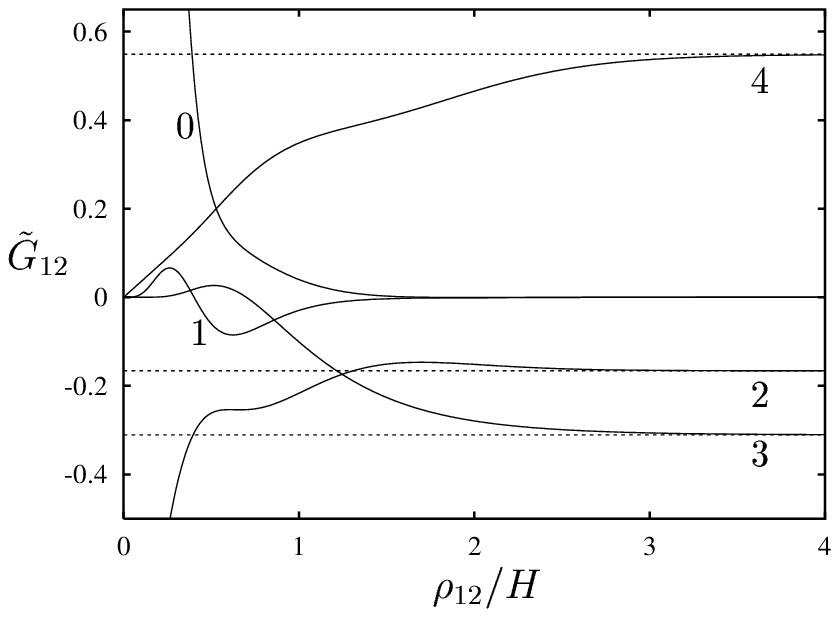}}}
\end{picture}
\end{center}
\caption{
Rescaled matrix elements
$\tilde\GreenWallTotElement_{12}(1\,\,\mbox{$-1$}\,0\mid 4\,m\,0)$
versus lateral distance $\LateralDistance_{12}$ for $Z_1=Z_2=\quarter
H$; values of parameter $m$ as labeled.  Exact solution (solid
lines); asymptotic behavior (dotted lines).
}
\label{plots of asymptotic behavior of matrix elements m}
\end{figure}

\begin{figure}
\begin{center}
\begin{picture}(300,360)(0,0)
\put(0,0){\scalebox{1}
      {\includegraphics{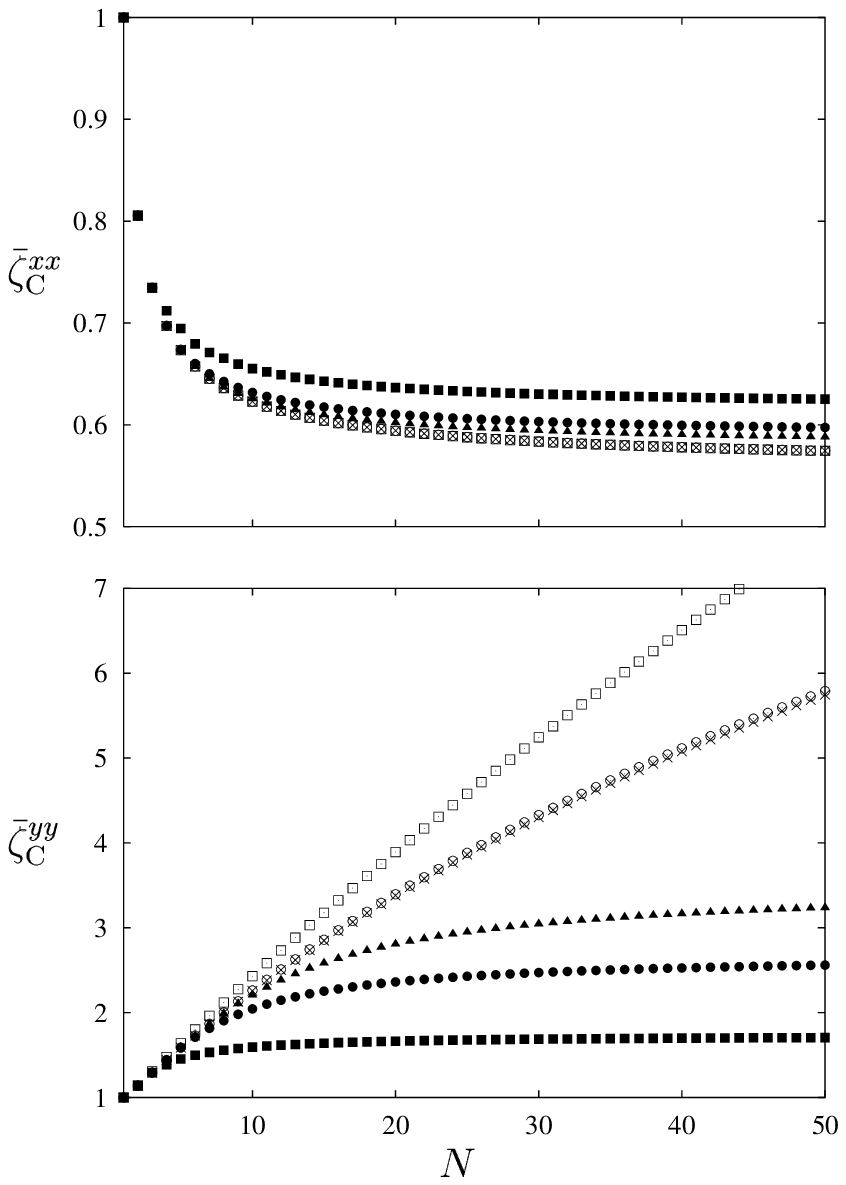}}}
\end{picture}
\end{center}
\caption{
Normalized longitudinal and transverse friction coefficients per
particle $\rigidBodyResistanceNorm^{xx}$ and
$\rigidBodyResistanceNorm^{yy}$ for linear arrays of touching spheres
in the center plane between the walls, for wall separation
$H/2a=1.05$.  Crosses represent exact results; open symbols represent
asymptotic approximation (\protect\ref{truncation of short-range
part}) with $\LateralDistance_\as/H=1$ (squares) and 2 (circles);
solid symbols correspond to truncation (\protect\ref{crude
truncation}) with $\LateralDistance_0/H=2$ (squares), 4 (circles) and 6
(triangles). For the longitudinal coefficient
$\rigidBodyResistanceNorm^{xx}$ the open squares and circles coincide
with the crosses.
}
\label{polymer}
\end{figure}

\begin{figure}
\begin{picture}(200,220)(0,0)
\put(0,0){\scalebox{.93}
      {\includegraphics{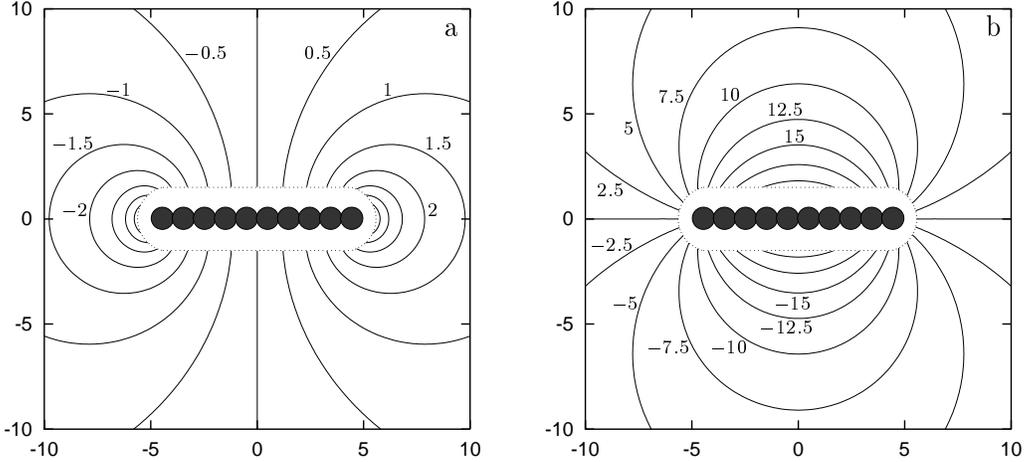}}}
\end{picture}
\caption{
Contour plots of the normalized asymptotic pressure field
(\protect\ref{normalized pressure}) around a rigid array of $N=10$
touching spheres moving in the center plane between parallel walls,
for wall separation $H/2a=1.05$.  Longitudinal motion \subfig{a};
transverse motion \subfig{b}.  The dotted lines delimit the regions
where the asymptotic approximation is not valid.
}
\label{pressure field 10}
\end{figure}

\begin{figure}[t]
\begin{picture}(200,220)(0,0)
\put(0,0){\scalebox{.93}
      {\includegraphics{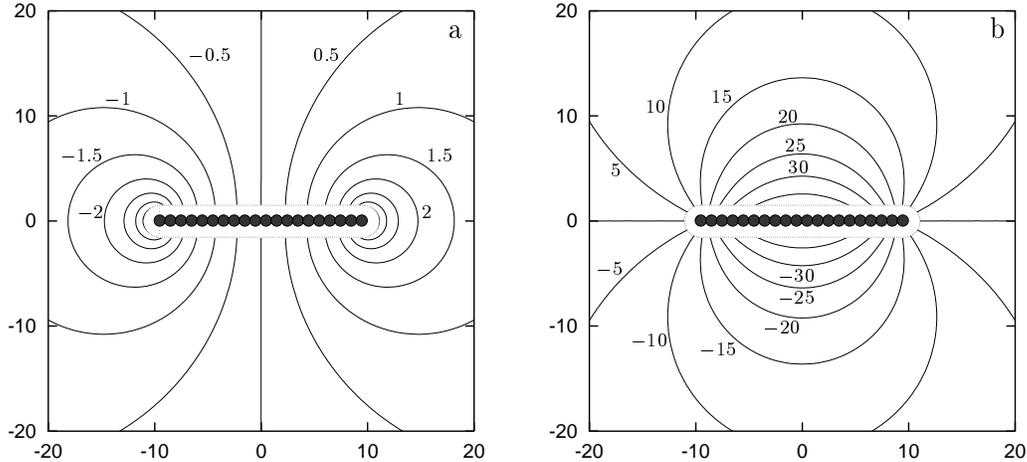}}}
\end{picture}
\caption{
Same as Fig.\ \protect\ref{pressure field 10}, except that for $N=20$.
}
\label{pressure field 20}
\end{figure}

Figure \ref{plots of asymptotic behavior of matrix elements 2}
illustrates the behavior of the matrix elements
\mbox{$\tilde\GreenWallTotElement_{12}(1\,\,\mbox{$-1$}\,0\mid
l\,1\,\sigma)$} with
\begin{equation}
\label{parameters of elements contributing to leading order}
l+\sigma\le3.
\end{equation}
All these functions approach nonzero asymptotic values
$\mbox{$\tilde\GreenWallTotElementAs_{12}(1\,\,\mbox{$-1$}\,0\mid
l\,1\,\sigma)$}\not=0$ for large interparticle distances
$\LateralDistance_{12}\gg H$, according to Eqs. \refeq{asymptotic
Green's matrix elements} and \refeq{condition for nonzero elements of
C matrix}.  The corresponding behavior of the unscaled matrix elements
\refeq{shown matrix elements}
\begin{equation}
\label{behavior of leading order elements}
\GreenWallTotElement_{12}
   (1\,\,\mbox{$\mp1$}\,0\mid l\,\,\mbox{$\pm1$}\,\sigma)
   \sim\LateralDistance_{12}^{-2},\qquad \LateralDistance_{12}\gg H,
\end{equation}
follows from Eqs.\ \refeq{Scalar basis minus}, \refeq{expression for
scalar displacement matrix}, and \refeq{asymptotic Green's matrix
elements}.  

The matrix elements \refeq{behavior of leading order elements} are
directly related to the multipolar flow fields \refeq{asymptotic wall
flow produced by force multipole}, as indicated by Eq.\
\refeq{asymptotic multipolar flow in terms of matrix elements}.
Therefore, we find that Eq.\ \refeq{behavior of leading order
elements} corresponds to the slowest possible far-field decay of the
flow produced by a multipolar force distribution.  The multipoles
\refeq{parameters of elements contributing to leading order} include
the horizontal Stokeslet ($l=1$, $\sigma=0$), rotlet ($l=1$,
$\sigma=1$), stresslet ($l=2$, $\sigma=0$), and three other
multipoles, one of which has the spherical-harmonics order $l=3$.  The
numerical results shown in Fig.\ \ref{plots of asymptotic behavior of
matrix elements 2} indicate that the approach of
$\tilde\GreenWallTotElement_{12}$ to the asymptotic values is
exponential, which is consistent with our discussion in Sec.\
\ref{Far-field approximation}.

Figure \ref{plots of asymptotic behavior of matrix elements m}
illustrates the behavior of matrix elements \refeq{shown matrix
elements} for $l=4$ and $\sigma=0$.  Unlike the elements presented in
Fig.\ \ref{plots of asymptotic behavior of matrix elements 2}, the
rescaled elements
\mbox{$\tilde\GreenWallTotElement_{12}(1\,\,\mbox{$-1$}\,0\mid
4\,m\,0)$} with $m=0,1$ vanish at large separations
$\LateralDistance\gg H$, which is consistent with condition
\refeq{condition for nonzero elements of C matrix}.  The remaining
rescaled elements with $m=2,3,4$ tend exponentially to nonzero
asymptotic values.

\subsection{Applications in multiparticle hydrodynamic-interactions 
algorithms}
\label{Applications in multiparticle hydrodynamic-interactions 
algorithms}

The simplest numerical application of our asymptotic formulas
\refeq{asymptotic Green's matrix elements}, \refeq{nonzero elements of
C matrix}, and \refeq{expression for matrix B} is to implement them
directly in the induced-force-multipole equation \refeq{induced force
equations in matrix notation}.  To this end, the matrix
\refeq{elements of total wall operator} is represented as the
superposition of the long-range asymptotic part and the short-range
correction, i.e.,
\begin{equation}
\label{Wall Green matrix as long range and short range parts}
\GreenWallTotElement_{ij}(lm\sigma\mid l'm'\sigma')
   =\GreenWallTotElementAs_{ij}(lm\sigma\mid l'm'\sigma')
   +\delta\GreenWallTotElement_{ij}(lm\sigma\mid l'm'\sigma').
\end{equation}
The asymptotic part \mbox{$\GreenWallTotElementAs_{ij}(lm\sigma\mid
l'm'\sigma')$} can be evaluated from our explicit formulas at a low
numerical cost.  To obtain the correction term
\mbox{$\delta\GreenWallTotElement_{ij}(lm\sigma\mid lm\sigma)$}, first
the expression \mbox{$\GreenWallTotElement_{ij}(lm\sigma\mid
l'm'\sigma')$} is calculated using the Cartesian-repre\-sen\-ta\-tion method
described in
\cite{Bhattacharya-Blawzdziewicz-Wajnryb:2005,%
Bhattacharya-Blawzdziewicz-Wajnryb:2005a%
}
and next, the asymptotic expression is subtracted from the result.
Since the correction term is short ranged, the matrix
$\delta\GreenWallTotElement_{ij}(lm\sigma\mid l'm'\sigma')$ can be
truncated by setting
\begin{equation}
\label{truncation of short-range part}
\delta\GreenWallTotElement_{ij}(lm\sigma\mid l'm'\sigma')=0
   \quad\mbox{for}\quad
   \LateralDistance_{ij}>\LateralDistance_\as,
\end{equation}
where the truncation distance $\LateralDistance_\as$ is of the order
of several wall separations $H$.  The results shown in Figs.\
\ref{plots of asymptotic behavior of matrix elements 2} and \ref{plots
of asymptotic behavior of matrix elements m} and other similar tests
indicate that the asymptotic approximation for the Green's matrix is
very accurate for $\LateralDistance_\as\gtrsim3H$.  Thus, the
numerically expensive contribution $\delta\GreenWallTotElement_{ij}$
has to be evaluated only for the neighboring particles at an $O(N)$
cost.

To test our asymptotic approach and illustrate the role of the long-
and short-range contributions to the Green's matrix
$\GreenWallTotElement$, we consider a benchmark case of a linear rigid
array of $N$ touching spheres translating in the center plane in the
space between closely spaced walls.  The spheres are on a line
parallel to the $x$ direction and the array is moving either in the
$x$ (longitudinal) or $y$ (transverse) direction.  We focus on the
translational friction coefficients evaluated per particle,
\begin{equation}
\label{rigid-body translational resistance}
\rigidBodyResistanceNorm^{\alpha\alpha}
   =(N\resistanceMatrixElement_\parallel)^{-1}
    \sum_{i,j=1}^N
      \resistanceMatrixElement_{ij}^{\transl\transl\,\alpha\alpha},
\qquad \alpha=x,y,
\end{equation}
where $\resistanceMatrixElement_{ij}^{\transl\transl\,\alpha\alpha}$
is the $\alpha\alpha$ component of the translational resistance tensor
$\resistanceMatrix_{ij}^{\transl\transl}$ defined in Eq.\
\refeq{resistance relation}, and $\resistanceMatrixElement_\parallel$
is the one-particle lateral translational resistance coefficient
\cite{Jones:2004,Bhattacharya-Blawzdziewicz-Wajnryb:2005}.

As illustrated in Fig.\ \ref{polymer} (see also discussion in
\cite{Bhattacharya-Blawzdziewicz-Wajnryb:2005a,%
Bhattacharya-Blawzdziewicz-Wajnryb:2005%
})
the longitudinal and transverse friction coefficients
\refeq{rigid-body translational resistance} behave differently.  The
longitudinal coefficient $\rigidBodyResistanceNorm^{xx}$ decreases
with the length of the array $N$, while the transverse coefficient
$\rigidBodyResistanceNorm^{yy}$ increases with $N$.  For tight
configurations with small gaps between the wall and particle surfaces
(the case show in Fig.\ \ref{polymer}) the decrease of
$\rigidBodyResistanceNorm^{xx}$ is moderate because the friction force
is dominated by the local resistance due to the dissipation in these
gaps.  In contrast, the increase of $\rigidBodyResistanceNorm^{yy}$ is
large due to collective long-range effects.

The mechanism of these collective effects can be explained using the
results for the pressure field around arrays of the length $N=10$ and
20 plotted in Figs.\ \ref{pressure field 10} and \ref{pressure field
20}.  The figures show the normalized asymptotic far-field pressure
\begin{equation}
\label{normalized pressure}
\bar\scatteredPressureAs=H(\eta U)^{-1}\scatteredPressureAs
\end{equation}
(where $U$ is the velocity of the array), which was evaluated using
the method described in Appendix \ref{Far-field pressure
distribution}.  The results for the longitudinal motion of the array
shown in Figs. \ref{pressure field 10}\subfig{a} and \ref{pressure
field 20}\subfig{a} indicate that the pressure field is only weakly
affected by the length of the array, and its magnitude is the largest
near the array ends.  In contrast, the pressure shown in
Figs. \ref{pressure field 10}\subfig{b} and \ref{pressure field
20}\subfig{b} for the transverse motion increases approximately
linearly with the array length $N$, and its magnitude is maximal near
the chain center.  This large pressure amplitude is associated with
the flow of the displaced fluid around the ends of the array in the
confined space.  The flow is significant over the distance that
scales with the length of the array $l=2Na$ (where $a$ is the sphere
radius).  In the Hele-Shaw regime the pressure gradient is
proportional to the fluid velocity; hence, the pressure itself is
proportional to $N$.

To further elucidate the effects of the short-range and far-field flow
components, the exact numerical results for the resistance coefficients
\refeq{rigid-body translational resistance} are compared in Fig.\
\ref{polymer} with the asymptotic approximation \refeq{Wall Green
matrix as long range and short range parts} and \refeq{truncation of
short-range part}.  We also show results obtained using a much cruder
approximation, where the {\it whole} Green's matrix is truncated at a
certain distance $\rho_0$, i.e.,
\begin{equation}
\label{crude truncation}
\GreenWallTotElement_{ij}(lm\sigma\mid l'm'\sigma')=0
   \quad\mbox{for}\quad
   \LateralDistance_{ij}>\LateralDistance_0.
\end{equation}
Our numerical calculations indicate that the truncation \refeq{crude
truncation} yields poor results.  The far-field flow contribution is
especially important for the transverse motion of the array because of
the positive-feedback effect: For this motion the dipolar Hele-Shaw
flow field generated by a given particle acts as a back flow on the
other particles.  This back flow, in turn, produces an increase in the
induced force distribution that generates the dipolar flow.  This back
flow mechanism, resulting in the large transverse resistance, is
consistent with our discussion of the pressure field shown in Figs.\
\ref{pressure field 10}\subfig{b} and \ref{pressure field
20}\subfig{b}.

In contrast to the crude approximation \refeq{crude truncation}, a
truncation of the short-range part \refeq{truncation of short-range
part} of the Green's matrix yields accurate results already with
moderate values of the truncation parameter $\LateralDistance_\as$.
The results shown in Fig.\ \ref{polymer} indicate that the truncations
at $\LateralDistance_\as/H=1$ for the longitudinal motion and at
$\LateralDistance_\as/H=2$ for the transverse motion are sufficient.
The results with $\LateralDistance_\as/H\ge3$ (not shown) are
essentially indistinguishable from the exact results.

\section{Conclusions}
\label{conclusions}

Our paper presents a complete analysis of the far-field flow produced by an
arbitrary force multipole in the space bounded by two parallel planar walls.
We have shown that a force multipole characterized by the multipolar numbers
$lm\sigma$ produces, at large distances, a Hele-Shaw flow driven by a
two-dimensional multipolar pressure field of the azimuthal order $m$.  The
amplitude of this flow has been explicitly obtained for an arbitrary order of
the source force multipole.

Our asymptotic results were applied to evaluate the multipolar matrix elements
\mbox{$\GreenWallTotElement_{ij}(lm\sigma\mid l'm'\sigma')$} of the Green's
tensor for Stokes flow in the wall-bounded domain.  This matrix is used in our
recently developed algorithm
\cite{
Bhattacharya-Blawzdziewicz-Wajnryb:2005,%
Bhattacharya-Blawzdziewicz-Wajnryb:2005a%
}
for evaluation of the multiparticle friction tensor $\resistanceMatrix_{ij}$
in a suspension confined between two parallel walls.  The elements of the
matrix $\GreenWallTotElement_{ij}$ are equivalent to the expansion
coefficients in the displacement theorem for Stokes flow in the bounded
domain.  Such a displacement theorem relates the flow produced by a force
multipole centered at a point $\bR_j$ to nonsingular multipolar flows centered
at a point $\bR_i$.  We have shown that in the far-field regime the matrix
elements \mbox{$\GreenWallTotElement_{ij}(lm\sigma\mid l'm'\sigma')$} can be
expressed in terms of much simpler displacement formulas for the
two-dimensional scalar potential.

We have found that the matrix $\GreenWallTotElement_{ij}$ achieves its
asymptotic behavior when the lateral distance between the centers of the
particles $i$ and $j$ exceeds several wall separations $H$.  Evaluation of the
exact matrix elements in terms of lateral Fourier integrals derived in
\cite{Bhattacharya-Blawzdziewicz-Wajnryb:2005a,%
Bhattacharya-Blawzdziewicz-Wajnryb:2005%
}
is thus needed only for the neighboring particles.  Therefore,
application of the asymptotic expressions in our
hydrodynamic-interaction algorithm yields an important improvement of
its numerical efficiency.  (The far-field contribution to the Green's
matrix cannot be simply neglected---for some problems such a crude
approximation leads to entirely wrong values of the friction matrix;
cf. discussion in section \ref{Applications in multiparticle
hydrodynamic-interactions algorithms}).

Several other important consequences stem from the fact that we have reduced a
complex hydrodynamic problem to a simpler problem of a two-dimensional scalar
potential.  First, since for a scalar potential the multipolar flow fields in
a periodic system are known \cite{Cichocki-Felderhof:1989a}, the results of
our analysis can be used to develop an algorithm for hydrodynamic interactions
in a periodic wall-bounded system.  Without the asymptotic expressions,
evaluation of the periodic hydrodynamic Green's matrix would be much more
difficult, as discussed in \cite{Bhattacharya:2005}.

Next, for scalar potentials, fast multipole and PPPM acceleration techniques
are well developed \cite{Frenkel-Smit:2002}.  Combined with our asymptotic
results, such methods can be applied for fast evaluation of hydrodynamic
interactions in wall-bounded suspensions.  Development of accelerated
algorithms for suspensions will require implementation of certain techniques
that are specific to multiparticle hydrodynamic systems, e.g. an appropriate
preconditioning of the Green's matrix and incorporating the lubrication
interactions into the calculation scheme.  These techniques were used in
accelerated Stokesian-dynamics algorithms for unbounded suspensions
\cite{Sierou-Brady:2001} and \cite{Sangani-Mo:1996}.  Our present asymptotic
results greatly facilitate development of accelerated algorithms for
wall-bounded systems, and our research is currently focused on this problem.

S.\,B.\ would like to acknowledge the support by NSF grant
CTS-0201131.  E.\,W.\ was supported by NASA grant NAG3-2704 and, in
part, by KBN grant No.\ 5T07C 035 22.  J.\,B.  was supported by NSF
grant CTS-S0348175.

\appendix

\section{Spherical basis}
\label{Spherical basis}

The spherical basis sets of Stokes flows $\sphericalBasisPM{lm\sigma}$
used in the present paper are normalized differently than the
corresponding sets $\bv^{\pm({\rm CFS})}_{lm\sigma}$
introduced in \cite{Cichocki-Felderhof-Schmitz:1988}.  The
transformations between the basis fields are
\begin{subequations}
\label{relation between BBW and CFS basis}
\begin{equation}
\label{relation between BBW and CFS v-}
\sphericalBasisM{lm\sigma}(\br)
   =N_{l\sigma}^{-1}n_{lm}^{-1}\bv^{-({\rm CFS})}_{lm\sigma}(\br),
\end{equation}
\begin{equation}
\label{relation between BBW and CFS v+}
\sphericalBasisP{lm\sigma}(\br)
   =N_{l\sigma}n_{lm}^{-1}\bv^{+({\rm CFS})}_{lm\sigma}(\br),
\end{equation}
\end{subequations}
where
\begin{subequations}
\begin{equation}
\label{change-of-normalization coefficients}
N_{l0}=1,
\end{equation}
\begin{equation}
N_{l1}=-{(l+1)^{-1}},
\end{equation}
\begin{equation}
N_{l2}=l[(l+1)(2l+1)(2l+3)]^{-1}, 
\end{equation}
\end{subequations}
and
\begin{equation}
\label{normalization coefficients}
n_{lm}=\left[\frac{4\pi}{2l+1} \frac{(l+m)!}{(l-m)!}\right]^{1/2}.
\end{equation} 

Below we list the explicit expressions for the angular coefficients
$\sphericalBasisCoefPM{lm\sigma}(\theta,\phi)$ for spherical basis
fields \refeq{spherical basis v +-} in our present normalization,
\begin{subequations}
\label{V-}
\begin{equation}
\label{V-0}
\sphericalBasisCoefM{lm0}=\frac{1}{(2l+1)^2}\left[
   \frac{l+1}{l(2l-1)}\alpha_l\bY_{l\,l-1\,m}
   -\frac{1}{2}\bY_{l\,l+1\,m}
\right],
\end{equation}
\begin{equation}
\label{V-1}
\sphericalBasisCoefM{lm1}=\frac{i}{l+1}\gamma_l\bY_{l\,l\,m},
\end{equation}
\begin{equation}
\label{V-2}
\sphericalBasisCoefM{lm2}=\beta_l\bY_{l\,l+1\,m},
\end{equation}
\end{subequations}
and
\begin{subequations}
\label{V+}
\begin{equation}
\label{V+0}
\sphericalBasisCoefP{lm0}=\alpha_l\bY_{l\,l-1\,m},
\end{equation}
\begin{equation}
\label{V+1}
\sphericalBasisCoefP{lm1}=\frac{\im}{l+1}\gamma_l\bY_{l\,l\,m},
\end{equation}
\begin{equation}
\label{V+2}
\sphericalBasisCoefP{lm2}=\frac{l}{2(2l+1)}\alpha_l\bY_{l\,l-1\,m}
   +\frac{l}{(l+1)(2l+1)(2l+3)}\beta_l\bY_{l\,l+1\,m},
\end{equation}
\end{subequations}
where
\begin{subequations}
\label{v-harmonics}
\begin{equation}
\label{vector harmonics A}
  {\bf Y}_{ll-1m}(\hr)=\alpha_l^{-1}
  r^{-l+1}\bnabla\left[r^l Y_{lm}(\hr)\right],
\end{equation}
\begin{equation}
\label{vector harmonics B}
  {\bf Y}_{ll+1m}(\hr)=\beta_l^{-1}
  r^{l+2}\bnabla\left[r^{-(l+1)} Y_{lm}(\hr)\right],
\end{equation}
\begin{equation}
\label{vector harmonics C}
  {\bf Y}_{llm}(\hr)=\gamma_l^{-1}\br\btimes\bnabla_{\mathrm{s}} Y_{lm}(\hr)
\end{equation}
\end{subequations}
are the  normalized
vector spherical harmonics, as defined by \cite{Edmonds:1960}.
Here
\begin{equation}
\label{scalar harmonics}
   Y_{lm}(\hr) =n_{lm}^{-1} (-1)^m P_l^m(\cos\theta)e^{{\rm i}m\varphi}
\end{equation}
are the normalized scalar spherical harmonics, and 
\begin{subequations}
\begin{equation}
\label{norm.const.}
  \alpha_l=[l(2l+1)]^{1/2},
\end{equation}
\begin{equation}
  \beta_l=[(l+1)(2l+1)]^{1/2},
\end{equation}
\begin{equation}
  \gamma_l=-\im[l(l+1)]^{1/2}.
\end{equation}
\end{subequations}

\section{Flow fields generated by force multipoles}
\label{Matrix elements and asymptotic flow}

In this Appendix we express the flow field
\begin{equation}
\label{multipolar flow fields between walls}
\wallSphericalBasisM{lm\sigma}(\br-\LateralDistance_2;Z_2)=
   \int \bT(\br,\br')
\deltab{a}(\br'-\bR_2)
      \reciprocalSphericalBasisP{lm\sigma}(\br'-\bR_2)\diff\br'
\end{equation}
produced in the space between the walls by the force multipole
\refeq{force multipole producing far field flow} in terms of the
elements of the Green's matrix $\GreenWallTotElement_{12}$.  Using
relation \refeq{expansion of Stokes flow in nonsingular basis} to
expand the right side of Eq.\ \refeq{multipolar flow fields between
walls} into the non-singular spherical basis fields \refeq{spherical
basis v +} yields
\begin{equation}
\label{expansion of wall multipolar fields into spherical + basis}
\wallSphericalBasisM{lm\sigma}(\br-\LateralVector_2;Z_2)
   =\sum_{l'm'\sigma'}\sphericalBasisP{l'm'\sigma'}(\br-\bR_1)
   \GreenWallTotElement_{12}(l'm'\sigma'\mid lm\sigma)
\end{equation}
where the definition \refeq{elements of total wall operator} of the
Green's matrix have been used.  Equation \refeq{expansion of
asymptotic multipolar fields into spherical + basis} is the asymptotic
form of the above relation.

Equation \refeq{expansion of wall multipolar fields into spherical +
basis} can be simplified by evaluating it for $\br=\bR_1$ and noting
that
\begin{equation}
\label{flow fields p vanishing at origin}
\sphericalBasisP{l'm'\sigma'}(0)=0\quad\mbox{for}\quad l'+\sigma'>1
\end{equation}
and
\begin{equation}
\label{spherical basis plus at 0}
\sphericalBasisP{1\,m'\,0}(0) =(\smallfrac{4}{3}\pi)^{-1/2}\hat\be_{m'},
\end{equation}
where $ m'=0,\pm1$, and
\begin{equation}
\label{complex unit vectors}
\hat\be_{\pm1}=\mp\frac{1}{\sqrt{2}}(\ex\pm\im\ey),\qquad\hat\be_0=\ez,
\end{equation}
which follows from Eqs.\ \refeq{V+}.  According to the above
expressions, only three terms of the sum \refeq{expansion of wall
multipolar fields into spherical + basis} contribute to the result,
\begin{equation}
\label{multipolar flow in terms of matrix elements}
\wallSphericalBasisM{lm\sigma}(\bR_1-\LateralVector_2;Z_2)
=(\smallfrac{4}{3}\pi)^{-1/2}
  \sum_{m'=-1,0,1}\GreenWallTotElement_{12}
     (1\,m'\,0\mid l\,m\,\sigma)\hat\be_{m'}.
\end{equation}

In the asymptotic regime $\LateralDistance_{12}\gg H$ this relation
simplifies because the asymptotic matrix elements \refeq{asymptotic
Green's matrix elements} vanish for $m'm\ge0$, as discussed in Sec.\
\ref{Far-field approximation}.  Taking this observation into account
we thus find
\begin{equation}
\label{asymptotic multipolar flow in terms of matrix elements}
\wallSphericalBasisAsM{l\,\,\pm\mu\,\sigma}(\bR_1-\LateralVector_2;Z_2)
   =(\smallfrac{4}{3}\pi)^{-1/2}\GreenWallTotElementAs_{12}
     (1\,\,\mbox{$\mp1$}\,\,0\mid l\,\,\mbox{$\pm\mu$}\,\,\sigma)
      \hat\be_{\mp 1}
\end{equation}
where $\mu=1,2,\ldots$ The dependence of the flow field
\refeq{asymptotic multipolar flow in terms of matrix elements} on the
lateral relative-position vector $\LateralVector_{12}$ can be made
more explicit by using relation \refeq{rescaled elements of Green
matrix} and the identity
\begin{equation}
\label{identity for derivative of scalar basis field}
\ScalarBasis{\pm(\mu+1)}(\LateralVector_{12})\hat\be_{\mp1}
  =\mp2^{-1/2}(\mu+1)^{-1}\bnablaLat\ScalarBasisM{\pm\mu}(\LateralVector_{12}),
\end{equation}
which yields
\begin{equation}
\label{form of far field flow}
\wallSphericalBasisAsM{l\,\pm\mu\sigma}(\bR_1-\LateralVector_2;Z_2)
   =\mp(\smallfrac{8}{3}\pi)^{-1/2}(\mu+1)^{-1}H^{\mu-l-\sigma+1}
   \tilde\GreenWallTotElementAs_{12}
            (1\,\,\mbox{$\mp1$}\,0\mid l\,\,\mbox{$\pm\mu$}\,\sigma)
   \bnablaLat\ScalarBasisM{\pm\mu}(\LateralVector_{12}).
\end{equation}
We note that relation \refeq{form of far field flow} is consistent
with \refeq{expression for multipolar asymptotic field in Hele-Shaw
basis} due to Eq.\ \refeq{Hele-Shaw basis velocity fields}.

\section{Far-field pressure distribution}
\label{Far-field pressure distribution}

As discussed in Sec.\ \ref{Far-field approximation}, the flow and the
pressure fields in the Hele-Shaw asymptotic regime \refeq{Hele-Shaw
flow field} are uniquely related (up to an additive constant in the
pressure).  Thus, many of the asymptotic formulas, expressed here in
terms of the velocity fields, can be translated into the corresponding
expressions for the pressure.

This remark applies, in particular, to Eq.\ \refeq{expression for
multipolar asymptotic field in Hele-Shaw basis} for the asymptotic
multipolar flow \refeq{asymptotic wall flow produced by force
multipole}.  We introduce the asymptotic multipolar pressure field
$\wallPressureAsM{lm\sigma}(\br)$, which is defined by the relation
\begin{equation}
\label{relation between multipolar pressure and velocity}
\wallSphericalBasisAsM{lm\sigma}(\br-\LateralVector_2;Z_2)
   =-\half\eta^{-1}z(H-z)
      \bnablaLat\wallPressureAsM{lm\sigma}
      (\lateralVector-\LateralVector_2;Z_2).
\end{equation}
Using Eqs.\ \refeq{pressure in terms of solid harmonics} and
\refeq{Hele-Shaw basis velocity fields}, the flow-field identity
\refeq{expression for multipolar asymptotic field in Hele-Shaw basis}
can be transformed into the corresponding pressure identity of the
form
\begin{equation}
\label{expression for multipolar asymptotic pressure in asymptotic  basis}
\wallPressureAsM{lm\sigma}(\lateralVector-\LateralVector_2;Z_2)
=-\frac{6}{\pi H^{3}}
\ScalarBasisM{m}(\lateralVector-\LateralVector_2)
\TransformationElementSAs(Z_2;lm\sigma).
\end{equation}

Equation \refeq{expression for multipolar asymptotic pressure in
asymptotic basis} can be conveniently used to evaluate the far-field
disturbance pressure $\scatteredPressureAs$ in a many-particle system.
This equation describes the asymptotic pressure produced in the
far-field regime by a single force multipole \refeq{force multipole
producing far field flow}, as indicated by Eqs. \refeq{asymptotic wall
flow produced by force multipole} and \refeq{relation between
multipolar pressure and velocity}.  To determine
$\scatteredPressureAs$, the multipolar moments $f_i(lm\sigma)$ of the
force distributions \refeq{induced force in terms of multipoles}
induced on the surfaces of particles $i=1,\ldots,N$ are evaluated by
solving the force-multipole equations \refeq{induced force equations
in matrix notation}.  Combining the solution with \refeq{expression
for multipolar asymptotic pressure in asymptotic basis} yields
\begin{equation}
\label{far field pressure in the system}
\scatteredPressureAs(\lateralVector)
=\sum_{i=1}^N\sumPrimA{m}{} Q_i(m) \ScalarBasisM{m}(\lateralVector-\LateralVector_i),
\end{equation}
where
\begin{equation}
\label{pressure multipolar moments}
Q_i(m)=-\frac{6}{\pi H^{3}}\sum_{l\sigma}
    \TransformationElementSAs(Z_i;lm\sigma)f_i(lm\sigma).
\end{equation}
The contour plots in Figs.\ \ref{pressure
field 10} and \ref{pressure field 20} were obtained using this method.

\bibliographystyle{unsrt} 
\bibliography{/home/jerzy/BIB/jbib}

\end{document}